\documentclass{SciPost}

\hypersetup{
    colorlinks,
    linkcolor={red!50!black},
    citecolor={blue!50!black},
    urlcolor={blue!80!black}
}

\usepackage[bitstream-charter]{mathdesign}
\DeclareSymbolFont{usualmathcal}{OMS}{cmsy}{m}{n}
\DeclareSymbolFontAlphabet{\mathcal}{usualmathcal}

\fancypagestyle{SPstyle}{
\fancyhf{}

\fancyfoot[C]{\textbf{\thepage}}
}

\usepackage{xspace}

\begin{document}

\pagestyle{SPstyle}

\begin{center}{\Large \textbf{\color{scipostdeepblue}{
Combined EFT interpretation of top quark, Higgs boson, electroweak and QCD measurements at CMS\\
}}}\end{center}

\begin{center}\textbf{
Niels Van den Bossche\textsuperscript{1$\star$} on behalf of the CMS Collaboration
}\end{center}

\begin{center}
{\bf 1} Ghent University, Ghent, Belgium
\\[\baselineskip]
$\star$ \href{mailto:email1}{\small \href{mailto:niels.vandenbossche@cern.ch}{niels.vandenbossche@cern.ch}}
\end{center}

\definecolor{palegray}{gray}{0.95}
\begin{center}
\colorbox{palegray}{
  \begin{tabular}{rr}
  \begin{minipage}{0.36\textwidth}
    \includegraphics[width=60mm,height=1.5cm]{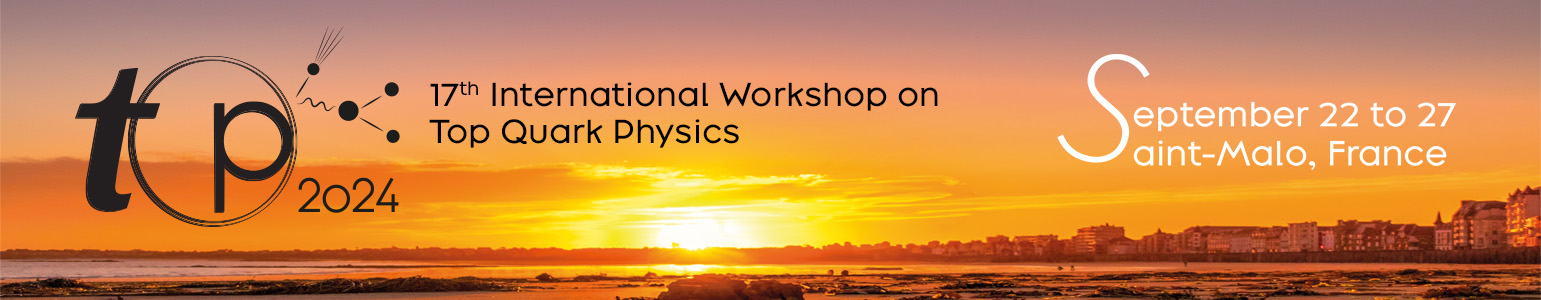}
  \end{minipage}
  &
  \begin{minipage}{0.55\textwidth}
    \begin{center} \hspace{5pt}
    {\it The 17th International Workshop on\\ Top Quark Physics (TOP2024)} \\
    {\it Saint-Malo, France, 22-27 September 2024
    }
    \end{center}
  \end{minipage}
\end{tabular}
}
\end{center}

\section*{\color{scipostdeepblue}{Abstract}}
\textbf{\boldmath{%
A recent EFT result from CMS is presented, combining differential cross section and direct EFT measurements
performed by the CMS Collaboration across four branches of the Standard Model: top, Higgs, electroweak and QCD physics.
To maximize the sensitivity, measurements of electroweak precision observables from LEP and SLC are included as well.
64 Wilson Coefficients (WC) are targeted in this combined measurement, both individually and 
with a simultaneous fit to 42 linear combinations of the Wilson Coefficients.
}}

\vspace{\baselineskip}





\newcommand{\mybar}[1]{\smash{$\bar{\text{#1}}$}}
\newcommand{\tbar}{\mybar{t}}
\newcommand{\ttbar}{t\tbar}
\newcommand{\tttt}{t\tbar{}t\tbar}
\newcommand{\ttW}{t\tbar W}
\newcommand{\ttZ}{t\tbar Z}
\newcommand{\ttH}{t\tbar H}

\section{Introduction}
\label{sec:intro}
Effective Field Theory (EFT) allows for the extension of the Standard Model (SM) Lagrangian with
various new physics effects in a model-independent way.
There are some implicit assumptions here, such as the new physics only arising at energies beyond the
reach of the experiments and respecting the SM gauge structure.
These extensions are additional terms in the Lagrangian, describing interactions 
with a higher mass dimension than present in the SM (dimension-5 and above)
and are expressed as a Wilson Coefficient (WC), acting as a coupling strength,
multiplied with an operator, which describes what fields interact with eachother in the extension.
Here, a specific SMEFT parameterization with 64 dimension-6 operators is considered 
using the topU3l basis of SMEFTSim3~\cite{smeftsim,smeftsimthree}.

The result discussed in these proceedings~\cite{CMS:SMP-24-003} is a reinterpretation and combination of 
various SM measurements at CMS~\cite{CMS_experiment,CMS_Run3}.
This result targets the tightest constraints on the SMEFT operators.
Seven previous results from CMS are combined, together with measurements 
of Electroweak Precision Observables (EWPO) from LEP and SLC.
As for CMS results, mostly differential cross section measurements are used.
Only one of the results is a direct EFT measurement, as will be discussed in the next section.

This result is not the first combination of this kind, but it is the first performed at CMS.
One of the main advantages of performing this type of combination within the collaboration 
is the availability of the full likelihoods of these results,
as well as the possibility to modify the existing analyses.
This also highlights the need to publish statistical models together with the papers 
from the collaborations.
While this has not been common practice, more results are now 
being published with the statistical model by the CMS collaboration.

\section{Input analyses}
The seven input analyses from CMS are selected to maximize the sensitivity 
to the considered operators, with the additional benefit 
of having a minimal overlap in event selection.

In the Higgs sector, a measurement of differential Higgs production cross sections across
all initial states is used as an input~\cite{CMS:HIG-19-015}.
As the final state, only the decay of the Higgs to two photons is considered.
The result is binned in the STXS stage 1.2 binning,
where the measured signal strength in each differential bin is parameterized 
for the EFT extraction as a function of the relevant operators.
To simulate the EFT effects, SMEFTSim3 is used where possible.
However, as some production processes contain loop diagrams,
SMEFT@NLO~\cite{smeftnlo} must be used.

In the electroweak sector, a set of three analyses targetting
$W\gamma$~\cite{CMS:SMP-20-005}, $WW$~\cite{CMS:SMP-18-004} and $Z\rightarrow\nu\nu$~\cite{CMS:SMP-18-003}
differential cross sections are used.
The first of the results is a double differential cross section measurement,
with the results in $p_T(\gamma)\times|\phi_f|$ used as the input to the EFT combination.
The other two results only use the data collected in 2016 and 
are single differential cross section measurements.
From the $WW$ measurement, the differential cross section of the invariant mass 
of the dilepton system is used, while for $Z\rightarrow\nu\nu$, the $p_T$ of the Z candidate is used.

In the QCD sector, a differential cross section measurement of inclusive jet production~\cite{CMS:SMP-20-011} is used.
From this result, the double differential cross section in $p_T$ and $\eta$ of AK7 jets is used.
AK7 jets specifically are chosen, as these are less sensitive to out-of-cone corrections.
A further modification to this result was made, by changing 
the PDF set in the original result from CT14 to CT18.

The most interesting inputs related to this conference are the inputs with top quark measurements.
Two results are used, one targetting \ttbar{} and one targetting all top processes with a multilepton final state.
The \ttbar{} result included is a differential cross section measurement in the single lepton final state~\cite{CMS:TOP-20-001}.
In this result, boosted top quarks are reconstructed, allowing to maximize the sensitivity
even at very high energies for the top quarks.
As an input distribution to the combined interpretation, the invariant mass of the \ttbar{}
pair is used.

\begin{figure}
  \centering
  \includegraphics[width=\textwidth]{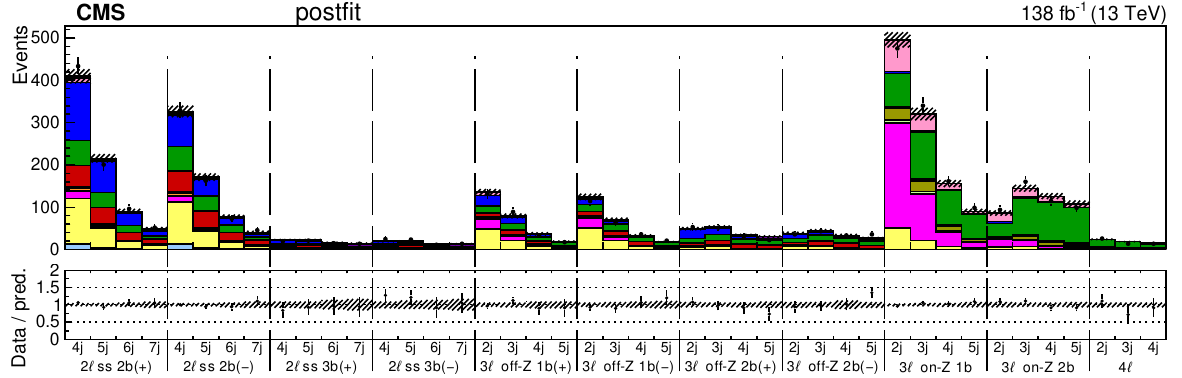}
  \caption{The binned events classification in the dedicated multilepton top EFT analysis~\cite{CMS:TOP-22-006}.}
  \label{fig:top22006regions}
\end{figure}

The last input, and the second result in the top physics sector, targets processes with top quarks
with a multilepton final state~\cite{CMS:TOP-22-006}.
The multilepton final state is defined as either a same-sign lepton pair, or three or more leptons.
The EFT effects on 6 processes are considered: \ttW, \ttZ, $tZq$, \ttH, $tHq$ and \tttt.
with off-shell contributions taken into account for the first three.
Events are binned based on lepton, b-jet and jet multiplicity, as well as the sum of the lepton charges.
In the three lepton final state, events are further split by the existence of a Z-boson candidate
in the event.
This Z-boson candidate is defined as an opposite-sign same-flavour 
lepton pair forming an invariant mass within 10 GeV of the mass of the Z boson.
This scheme defines in total 43 regions, shown in Figure~\ref{fig:top22006regions}.

The input to the EFT fit does not use the inclusive regions as shown in that figure, 
rather further binning the events in each region using a kinematic observable.
The variable used is the maximal $p_T$ of any combination of two leptons, jets or leptons and jets.
In the three-lepton final states with a Z boson candidate, the $p_T$ of this Z candidate is used.

\section{Combined EFT interpretation}
\begin{figure}
    \centering
    \includegraphics[width=\textwidth]{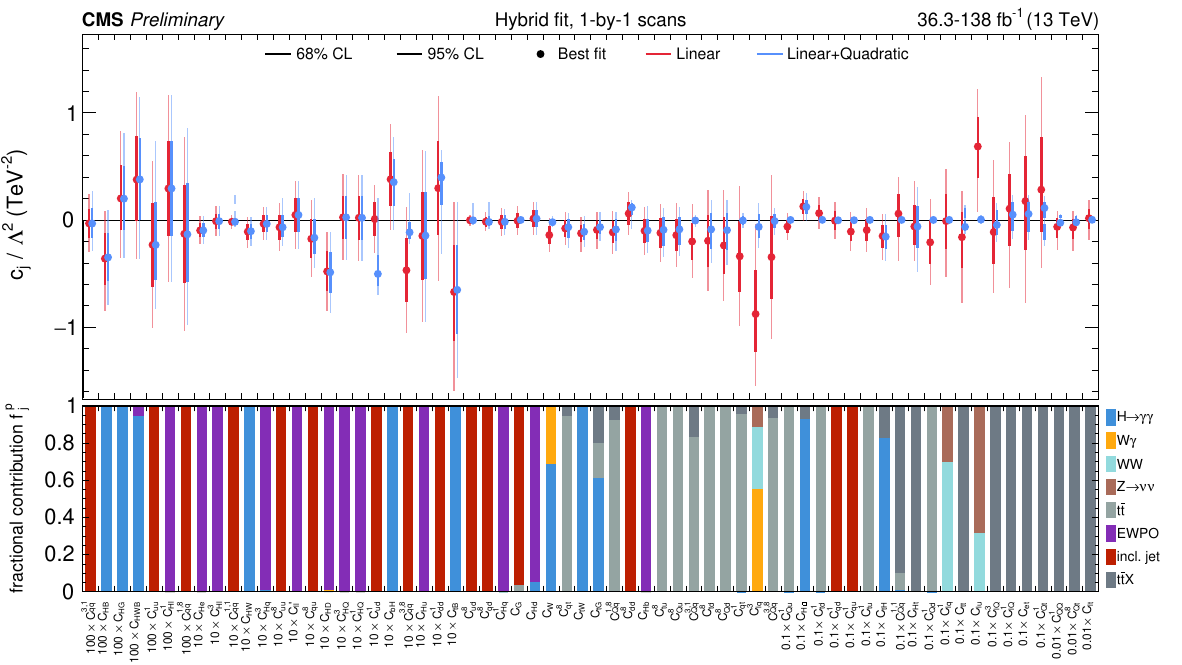}
    \caption{The obtained limits on individual WCs, with and without taking into account the 
    EFT contribution quadratic in the WC~\cite{CMS:SMP-24-003}.}
    \label{fig:EFT_LimitsLinQuad}
\end{figure}

Two results are reported.
First is the sensitivity to the individual WCs,
fitting only one at a time, while keeping the other WCs at their SM value of 0.
The results of these fits are shown in Figure~\ref{fig:EFT_LimitsLinQuad}.
Two confidence intervals are shown for each WC:
one where only contributions linear in the WC are included 
(i.e. the interference effects between SM diagrams and EFT diagrams),
and one with both linear and quadratic contributions in the WCs.
In the latter case, it has been verified that the coverage of the 
obtained confidence intervals is in reasonable agreement with the
target coverage using pseudodata samples to construct the test-statistic distributions.
Lastly, the lower panel of the figure highlights the contribution 
each included analysis has on the limit of each WC.

\begin{figure}
    \centering
    \includegraphics[width=\textwidth]{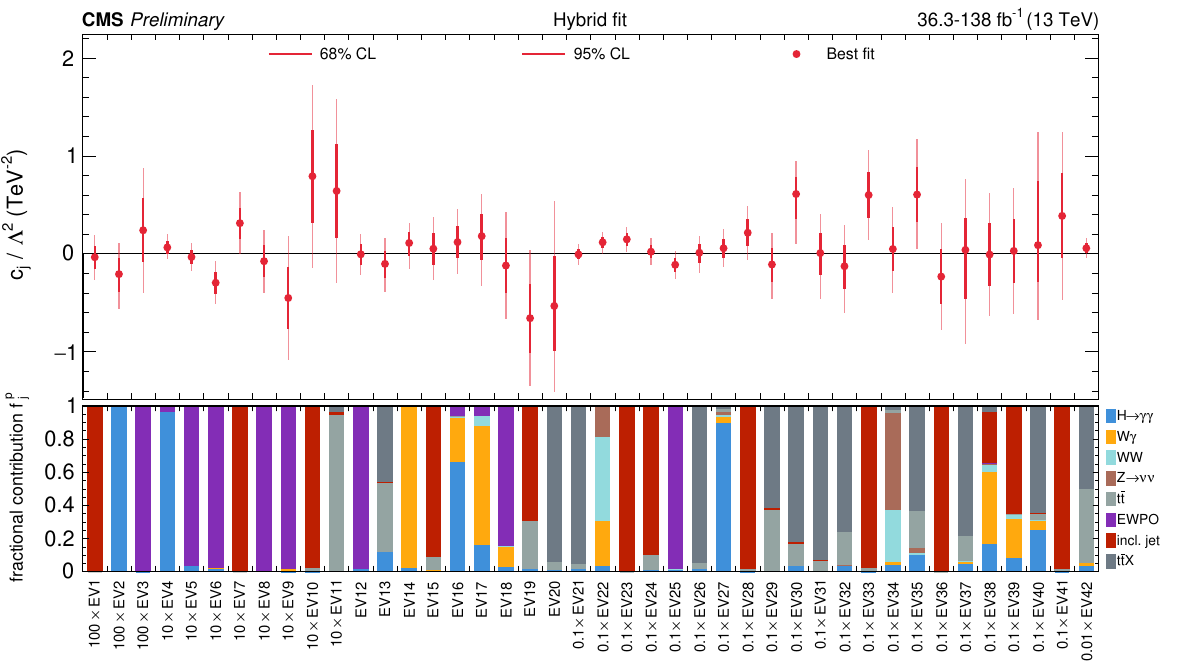}
    \caption{The obtained limits on the various linear combinations of WCs. 
    Only the contribution linear in the WC are considered~\cite{CMS:SMP-24-003}.}
    \label{fig:EFT_LimitsLinComb}
\end{figure}

The second result is a simultaneous fit of all WCs.
However, due to degeneracies, where different WCs affect
processes in the same way, not all of them can be constrained.
To account for this, a principal component analysis (PCA) is performed of
the Hessian matrix of the fit by performing a diagonalization. 
In this procedure, the likelihood function is constructed with only the components
linear in the WCs of the EFT parameterization.
By diagonalizing this Hessian matrix, an uncorrelated
set of linear combinations of all WCs can be obtained (the principal components).
As these are the eigenvectors of the Hessian matrix, they also contain an eigenvalue.
This eigenvalue is a measure for how strong the limits 
on the Wilson Coefficients can be obtained.
In the final result, only the linear combinations with an eigenvalue larger than 0.04
are considered, as those will have a sufficiently strong limit.
The remaining linear combinations remain for now unconstrained.
The fit result is shown in Figure~\ref{fig:EFT_LimitsLinComb}.

\section{Conclusion}
A recent result from CMS was presented, targetting 64 WCs of the SMEFT.
Two results are obtained, one with each WC fitted individually,
and one with 42 linear combinations of the considered WCs fitted simultaneously.
No significant deviations from the SM predictions have been observed.
In addition, special care has been taken to ensure the applicability of the obtained
confidence intervals as presented in the original paper~\cite{CMS:SMP-24-003},
even if the contributions quadratic in the WC are considered.





\begin{thebibliography}{99}
\newcommand{\arxiv}[1]{\href{http://arxiv.org/abs/#1}{arXiv:#1}}
\newcommand{\pas}[1]{\href{http://cms-results.web.cern.ch/cms-results/public-results/preliminary-results/#1/index.html}{CMS-PAS-#1}}
\newcommand{\etal}{{et al.}}
\newcommand{\titl}[1]{\textit{#1},\ }
\newcommand{\jrnl}[1]{\textit{#1}\xspace}
\newcommand{\vlme}[1]{\textbf{#1}\xspace}
\setlength{\itemsep}{-0.5\parskip}

\bibitem{smeftsim}
I. Brivio, Y. Jiang \& M. Trott,
\titl{The SMEFTsim package, theory and tools}%
\jrnl{JHEP} \vlme{12} (2017) 070, \doi{10.1007/JHEP12(2017)070}.

\bibitem{smeftsimthree}
I. Brivio,
\titl{SMEFTsim 3.0 -- a practical guide}%
\jrnl{JHEP} \vlme{04} (2021) 073, \doi{10.1007/JHEP04(2021)073}.

\bibitem{CMS:SMP-24-003}
CMS Collaboration,
\titl{Combined effective field theory interpretation of Higgs boson, electroweak vector boson, top quark, and multi-jet measurements}%
\arxiv{2504.02958}.

\bibitem{CMS_experiment}
CMS Collaboration,
\titl{The CMS experiment at the CERN LHC}%
\jrnl{JINST} \vlme{3} (2008) S08004, \doi{10.1088/1748-0221/3/08/S08004}.

\bibitem{CMS_Run3}
CMS Collaboration,
\titl{Development of the CMS detector for the CERN LHC Run 3}%
\jrnl{JINST} \vlme{19} (2024) 05, P05064, \doi{10.1088/1748-0221/19/05/P05064}.

\bibitem{CMS:HIG-19-015}
CMS Collaboration,
\titl{Measurements of Higgs boson production cross sections and couplings in the diphoton decay channel at $\sqrt{s}=13$ TeV}%
\jrnl{JHEP} \vlme{07} (2021) 027, \doi{10.1007/JHEP07(2021)027}

\bibitem{smeftnlo}
C. Degrande \etal,
\titl{Automated one-loop computations in the SMEFT}%
\jrnl{Phys. Rev. D} \vlme{103} (2021) 096024, \doi{10.1103/PhysRevD.103.096024}.

\bibitem{CMS:SMP-20-005}
CMS Collaboration,
\titl{Measurement of $W^\pm\gamma$ differential cross sections in proton-proton collisions at $\sqrt{s}=13$ TeV and effective field theory constraints}%
\jrnl{Phys. Rev. D} \vlme{105} (2022) 052003, \doi{10.1103/PhysRevD.105.052003}

\bibitem{CMS:SMP-18-004}
CMS Collaboration,
\titl{$W^+W^-$ boson pair production in proton-proton collisions at $\sqrt{s}=13$ TeV}
\jrnl{Phys. Rev. D} \vlme{102} (2020) 092001, \doi{10.1103/PhysRevD.102.092001}

\bibitem{CMS:SMP-18-003}
CMS Collaboration,
\titl{Measurement of the Z boson differential production cross section using its invisible decay mode ($Z\rightarrow \nu\bar{\nu}$) in proton-proton collisions at $\sqrt{s}=13$ TeV}%
\jrnl{JHEP} \vlme{05} (2021) 205, \doi{10.1007/JHEP05(2021)205}

\bibitem{CMS:SMP-20-011}
CMS Collaboration,
\titl{Measurement and QCD analysis of double-differential inclusive jet cross sections in proton-proton collisions at $\sqrt{s}=13$ TeV}%
\jrnl{JHEP} \vlme{02} (2022) 142, \doi{10.1007/JHEP02(2022)142}

\bibitem{CMS:TOP-20-001}
CMS Collaboration,
\titl{Measurement of differential \ttbar{} production cross sections in the full kinematic range using lepton + jets events from proton-proton collision at $\sqrt{s}=13$ TeV}%
\jrnl{Phys. Rev. D} \vlme{104} (2021) 092013, \doi{10.1103/PhysRevD.104.092013}

\bibitem{CMS:TOP-22-006}
CMS Collaboration,
\titl{Search for physics beyond the standard model in top quark production with additional leptons in the context of effective field theory}%
\jrnl{JHEP} \vlme{12} (2023) 068, \doi{10.1007/JHEP12(2023)068}

\end{thebibliography}
\end{document}